\documentclass{elsart}

\usepackage{graphicx}
\usepackage[english]{babel}
\usepackage{latexsym}
\usepackage{epsfig}
\usepackage{amssymb}
\usepackage[linktocpage=true,backref=true,breaklinks=true,colorlinks=true,linkcolor=blue,urlcolor=blue,citecolor=blue,anchorcolor=blue]{hyperref}

\begin{document} 

\begin{frontmatter}

\title{Pseudo-Newtonian Planar Circular Restricted 3-Body Problem}

\author{F. L. Dubeibe}
\ead{fldubeibem@unal.edu.co}

\address{Facultad de Ciencias Humanas y de la Educaci\'on, Universidad de los Llanos, 
Villavicencio, Colombia and \\
Grupo de Investigaci\'on en Relatividad y Gravitaci\'on, Escuela de F\'isica, Universidad Industrial de Santander, A.A. 678, Bucaramanga, Colombia}

\author{F. D. Lora-Clavijo \and Guillermo A. Gonz\'alez}
\ead{fadulora@uis.edu.co}
\ead{guillermo.gonzalez@saber.uis.edu.co}

\address{Grupo de Investigaci\'on en Relatividad y Gravitaci\'on, Escuela de F\'isica, Universidad Industrial de Santander, A.A. 678, Bucaramanga, Colombia}


\begin{abstract}
We study the dynamics of the planar circular restricted three-body problem in the context of a pseudo-Newtonian approximation. By using the Fodor-Hoenselaers-Perj\'es procedure, we perform an expansion in the mass potential of a static massive spherical source up to the first non-Newtonian term, giving place to a gravitational potential that includes first-order general relativistic effects. With this result, we model a system composed by two pseudo-Newtonian primaries describing circular orbits around their common center of mass, and a test particle orbiting the system in the equatorial plane. The dynamics of the new system of equations is studied in terms of the Poincar\'e section method and the Lyapunov exponents, where the introduction of a new parameter $\epsilon$, allows us to observe the transition from the Newtonian to the pseudo-Newtonian regime. We show that when the Jacobian constant is fixed, a chaotic orbit in the Newtonian regime can be either chaotic or regular in the pseudo-Newtonian approach. As a general result, we find that most of the pseudo-Newtonian configurations are less stable than their Newtonian equivalent.
\end{abstract}

\end{frontmatter}

\section{Introduction}

One of the simplest and most frequently studied version of the general three-body problem, is the planar circular restricted three-body problem (henceforth CRTBP), which can be stated as follows:
\begin{itemize}
    \item Two primaries, $\mathcal{M}_1$ and $\mathcal{M}_2$ at positions $X_1$ and $X_2$, respectively, follow a circular orbit around their common center of mass keeping a fixed distance $r$, while moving at constant angular velocity $\omega_0$.
    \item A third body $\mathcal{M}$, that is much smaller than either $\mathcal{M}_1$ or $\mathcal{M}_2$, remains in the orbital plane of the primaries.
    \item The equations of motion are derived only for the test particle $\mathcal{M}$, whose motion does not affect the primaries. 
\end{itemize}
The basic formulation of the CRTBP dates back to Euler, who proposed the use of synodical coordinates $(x,y)$ instead of the inertial coordinate system $(X,Y)$, in order to simplify this problem \cite{Euler67}. The transformation between these two systems can be performed by means of the rotation matrix\footnote{Along the paper $G=\mathcal{M}=\omega_{0}=r=1$, is understood.}, 

\begin{equation}\label{eq:rotmatrix}
 \left(  \begin{array}{c} x \\ y \end{array}\right)= 
 \left(   \begin{array} {cc}
  \cos\omega_0 t  & -\sin\omega_0 t  \\ \sin\omega_0 t & \cos\omega_0 t  
  \end{array}  \right) 
  \left( \begin{array}{c} X \\ Y \end{array} \right).
\end{equation}
Using this transformation, Lagrange proved the existence of five equilibrium points for the system, named Lagrangian points. The subject of  equilibrium points in the CRTBP has been studied extensively in the literature (see {\it e.g} \cite{Henon-book} and references therein). The discovery of the Trojan asteroids around the Lagrangian points $L_4$ and $L_5$ in the Sun-Jupiter system \cite{Murray-book}, and the recent observations of asteroids around $L_4$ for the Sun-Earth system \cite{Connors2011}, increased theoretical research on the subject (see {\it e.g.} \cite{Morbidelli2005}).  It should be noted that, in spite of the fact that the CRTBP is much simpler than the general three-body problem, it is non-integrable, which opened the possibility to analyze systematically the orbits \cite{Henon-book,Stephani-Book}.

Under the assumption of weak fields and low velocities, and as a first approximation to the relativistic CRTBP, in 1967 \cite{Krefetz1967} Krefetz considered for the first time the post-Newtonian equations of motion for the CRTBP, using the Einstein-Infeld-Hoffmann (EIH) formalism \cite{Einstein1937}.  The Lagrangian for this system was explicitly presented by Contopoulos in 1976 \cite{Contopoulos1976} and some typos for the Jacobian constant were corrected by Maindl \etal \cite{Maindl1994}, who also studied the deviations due to the post-Newtonian corrections on the Lagrangian points \cite{Maindl1996}. Concerning analytical solutions to the general relativistic three-body problem, Yamada \etal \cite{Yamada2010} obtained a  collinear solution by using the EIH approximation up to the first order. In a later paper, they studied the post-Newtonian triangular solution for three finite masses, showing that such configuration is not always equilateral \cite{Yamada2012}. Recently, as a first study of chaos in the post-Newtonian CRTBP, Huang and Wu \cite{Huang2014} studied the influence of the separation between the primaries, concluding that if it is close enough, the post-Newtonian dynamics is qualitatively different. In particular, some Newtonian bounded orbits become unstable.

To avoid the cumbersome equations of motion that take place in the post-Newtonian formalism, Steklain and Letelier used the Paczy\'nski--Wiita pseudo-Newtonian potential to study the dynamics of the CRTBP in the Hill's approximation \cite{Steklain2006}, finding that some pseudo-Newtonian systems are more stable than their Newtonian counterparts. Following this idea, and considering that the Jacobian constant is not preserved in the post-Newtonian approximation (which limits the dynamical studies), in the present paper we shall use an alternative approach to studying the dynamics of the pseudo-Newtonian CRTBP. To do so, we derive an approximate potential for the gravitational field of two uncharged spinless particles modeled as sources with multipole moment $m$, by using the Fodor-Hoenselaers-Perj\'es  (FHP)  procedure \cite{Fodor1989} (taking into account the corrections made by Sotiriou and Apostolatos \cite{Sotiriou2004}). Abusing astrophysical terminology, we call the new potential pseudo-Newtonian, due to the fact that in this kind of approaches the common Newtonian formulas are used even when the resulting potentials do not satisfy the Laplace equation. Unlike other pseudo-Newtonian approaches, the final expressions are not ad-hoc proposals but are derived directly from the multipole structure of the source. 

The paper is organized as follows: In section \ref{sec:PNMEEM}, by means of the FHP procedure, we calculate the  gravitational pseudo-potential for each primary; then we write down the Lagrangian of the CRTBP with their respective equations of motion for a test particle under the influence of this potential. In section \ref{sec:APND},  we analyze the gradual transition of the dynamics for the FHP pseudo-Newtonian approximation to the classical regime. The analysis is made using Poincar\'e surfaces of section and the variational method for the calculation of the largest Lyapunov exponent \cite{Contopoulos1978}.  Finally, in section \ref{sec:CR} we summarize our main conclusions.


\section{Pseudo-Newtonian Equations of Motion}
\label{sec:PNMEEM}

The Fodor-Hoenselaers-Perj\'es procedure is an algorithm to determine the multipole moments of stationary axisymmetric electrovacuum space-times \cite{Fodor1989}. The method can be stated as follows: 
In the Ernst formalism \cite{Ernst1968-1,Ernst1968-2}, Einstein field equations are reduced to a pair of complex equations through the introduction of the complex potentials  $\mathcal{E}$ and $\Psi$, which can be defined in terms of the new potentials $\xi$ and $\varsigma$, through the definitions
\begin{equation}
\mathcal{E}= \frac{1-\xi}{1+\xi}, \quad \Psi=\frac{\varsigma}{1+\xi},
\end{equation}
satisfying the alternative representation of the Einstein-Maxwell field
equations,
\begin{eqnarray}
\label{eq:GenerLaplaceXI}
(\xi \xi^{*} - \varsigma \varsigma^{*} -1)\nabla^2\xi =
2( \xi^{*} \nabla\xi -  \varsigma^{*} \nabla\varsigma)\cdot \nabla\xi,
\\
\label{eq:GenerLaplaceSIGMA}
(\xi \xi^{*} - \varsigma \varsigma^{*} -1)\nabla^2\varsigma =
2( \xi^{*} \nabla\xi -  \varsigma^{*} \nabla\varsigma)\cdot \nabla\varsigma.
\end{eqnarray}
The fields  $\xi$ and $\varsigma$ are related to the gravitational and electromagnetic potentials in a very direct way, 

\begin{equation}
\xi=\phi_{M}+i \phi_{J}, \quad \varsigma=\phi_{E}+i \phi_{H},
\end{equation}
where $\phi_{\alpha}$ with $\alpha=M,J,E,H$, are analogous to the Newtonian mass, angular momentum, electrostatic and magnetic potentials, respectively (see {\it e.g.} \cite{Hansen1974} and \cite{Sotiriou2004}). Hereafter, for the sake of convenience, we consider $\phi_{E}=\phi_{H}=0$, which implies $\varsigma=\Psi=0$, {\it i.e} the absence of electromagnetic fields. 

Now, according to Geroch and Hansen \cite{Hansen1974,Geroch1970}, the multipole moments of a given space-time are defined by measuring the deviation from flatness at infinity. Following this idea, the initial 3-metric $h_{ij}$ is mapped to a conformal one, that is $h_{ij} \rightarrow
\tilde{h_{ij}} = \Omega^{2}h_{ij}$. The conformal factor $\Omega$ transforms the potential $\xi$ into $\tilde{\xi} = \Omega^{-1/2}\xi$, with 
$\Omega = \bar{r}^{2} = \bar{\rho}^{2} + \bar{z}^{2}$, and
\begin{eqnarray}\label{eq:coordtrans}
\bar{\rho}&=&\frac{\rho}{\rho^{2}+z^{2}},\quad \bar{z}=\frac{z}{\rho^{2}+z^{2}}, \quad \bar{\varphi}=\varphi.
\end{eqnarray}
On the other hand, the potential $\tilde{\xi}$ can be written in a power series of  $\bar{\rho}$ and
$\bar{z}$ as
\begin{equation}
\label{eq:xi-q}
\tilde{\xi} = \sum_{i,j=0}^{\infty} a_{ij}\bar{\rho}^{i}\bar{z}^{j},    
\end{equation}
and the coefficients $a_{ij}$ are calculated by the
recursive relations \cite{Sotiriou2004}
\begin{eqnarray}\label{eq:rra}
(r + 2)^{2} a_{r+2,s} &=& -(s + 2)(s + 1)a_{r,s+2} + \sum_{k,l,m,n,p,g}(a_{kl}a^{*}_{mn}
- b_{kl}b^{*}_{mn}) [a_{pg}\nonumber\\
&\times& (p^{2} + g^{2} - 4p - 5g - 2pk  - 2gl- 2) + a_{p+2,g-2}(p + 2)\nonumber\\
&\times&(p + 2 - 2k)+ a_{p-2,g+2}(g + 2)(g + 1 - 2l)],
\end{eqnarray}
where $m = r - k - p, 0 \leq k \leq r, 0 \leq p \leq r - k$, with $k$ and $p$ even, and $n = s - l - g$,
$0 \leq l \leq s + 1$, and $-1 \leq g \leq s - l$.  Finally, the gravitational multipole moments $P_i$ of the source are computed from their values on the symmetry axis $m_{i}\equiv a_{0i}$, by means of the following relationships 
\begin{eqnarray}\label{eq:gpot}
P_0&=&m_0,\quad P_1=m_1,\quad P_2=m_2,\quad P_3=m_3,\quad P_4=m_4-\frac{1}{7}m_0^{*}M_{20},\nonumber\\
P_5&=&m_5-\frac{1}{3} m_{0}^{*}M_{30}-\frac{1}{21} m_{1}^{*}M_{20} 
\end{eqnarray}
where $M_{ij}=m_i m_j-m_{i-1}m_{j+1}$.

From the previous, it can be inferred that once we know the gravitational multipole moments, and following the inverse procedure, it is possible to determinate approximate expressions for the gravitational potential $\xi$,  in terms of the physical parameters of the source (see {\it e.g.} \cite{Pachon2011}).  Hence, let us apply the outlined procedure to a concrete example, a source whose multipole structure is given by 
\begin{equation}
P_{0}=m, P_{i}=0\quad {\rm for}\quad i\geq 1, 
\end{equation}
such that $m$ denotes the mass of the source. From Eq. (\ref{eq:rra}) and the seed $a_{00}=m$, it can be noted that the only non-vanishing coefficients are $a_{2n,2m}$ with $n, m \in \mathbb{N}$. Thus the potential 
$\xi$ is reconstructed from (\ref{eq:xi-q}),  (\ref{eq:coordtrans}) and  $ \xi= \Omega^{1/2}\tilde{\xi}$, and is explicitly given by
\begin{equation}\label{xiapp}
\xi(\rho,z)= \frac{m}{\sqrt{\rho^2+z^2}}-\frac{m^3 \rho^2}{2 \left(\rho^2+z^2\right)^{5/2}}+\frac{m^5 \rho^2 \left(3 \rho^2-4 z^2\right)}{8 \left(\rho^2+z^2\right)^{9/2}}+\ldots
\end{equation} 
It is important to note that the infinite sum of terms of the potential $\xi$ corresponds to the Schwarzschild solution, while the lower order of approximation, {\it i.e} by taking the first term, corresponds to the Newtonian potential. In order to stress that the approximations different to the lower order do not satisfy the Laplace equation, we call the potential $\xi=\phi_{M}$, pseudo-Newtonian.  

On the other hand, aiming to set up the planar circular restricted three-body problem in our model, we made the following assumptions:
{\it i)} The primaries are sufficiently far apart to keep moving in a circle; {\it ii)}  the superposition principle holds, such that the total gravitational potential $V$ can be expressed as $V=\phi_{M1}+\phi_{M2}$; {\it iii)} we consider only the first correction to the Newtonian potential, and 
{\it iv)} the motion takes place in the plane $z=0$.
Accordingly, the total potential energy of a test particle with mass $\mathcal{M}=1$ in the presence of two pseudo-Newtonian sources, in Cartesian coordinates, can be written as \footnote{The speed of light $c$ is explicitly presented in the pseudo-Newtonian potential in order to show its contributions; however, we set $c = 1$ in the numerical simulations.}
\begin{eqnarray}
V(x,y)  &=& -\sum_{i=1}^2\frac{\mathcal{M}_i}{r_i} +\frac{1}{2 c^4}\sum_{i=1}^2\frac{\mathcal{M}_i^3}{r_i^3} \label{eq:Uf}
\end{eqnarray} 
where $\mathcal{M}_{1}$ and $\mathcal{M}_{2}$ are the masses of each primary at positions $(x_1,0)$ and $(x_2,0)$, respectively, and $r_{1,2}=\sqrt{(x-x_{1,2})^2+y^2}$. Thus, the Lagrangian for the test particle moving in a non-inertial rotating frame, whose origin coincides with the center of mass of the system, in the presence of the potential (\ref{eq:Uf}) is expressed as
\begin{equation}\label{eq:LagF}
\mathcal{L}= \frac{U^2+2 A+R^2}{2} +\sum_{i=1}^2\frac{\mathcal{M}_i}{r_i}-\frac{1}{2c^4}\sum_{i=1}^2\frac{\mathcal{M}_i^3}{r_i^3},
\end{equation}
with $R=(x^2+y^2)^{1/2}$ the position of the test particle with respect to the center of mass, $U=(U_x^2+U_y^2)^{1/2}$ the magnitude of the velocity of the test particle in the rotating frame and $A=U_y x-U_x y$. Consequently, the Euler-Lagrange equations of motion are 
\begin{eqnarray}
&\ddot{x}=2 U_{y} + x - \sum_{i=1}^2\frac{\mathcal{M}_{i}(x-x_{i})}{r_{i}^3}
+ \frac{3}{2c^4}
\frac{\mathcal{M}_{i}^3(x-x_{i})}{r_{i}^5},\label{ecFHPx}\,\,\\
&\ddot{y}=-2 U_{x}+y \bigg(1-\sum_{i=1}^2\frac{\mathcal{M}_{i}}{r_{i}^3}+ \frac{3}{2c^4}\frac{\mathcal{M}_{i}^3}{r_{i}^5}\bigg).\label{ecFHPy}
\end{eqnarray}

Finally, the Jacobian integral of motion, for this approximation, can be calculated as
\begin{equation}\label{eq:JacF}
C_{J}= R^2-U^2+ 2 \sum_{i=1}^2\frac{\mathcal{M}_i}{r_i}-\frac{1}{c^4}\sum_{i=1}^2\frac{\mathcal{M}_i^3}{r_i^3}.
\end{equation}
It can be seen that, the equations (\ref{eq:LagF})-(\ref{eq:JacF}) reduce to the Newtonian case in the limit $1/c \to 0$. With some straightforward algebra, it can be easily shown that the Jacobian constan is conserved, that is to say, $d C_{J}/dt=0$.

\section{Analysis of the pseudo-Newtonian dynamics}\label{sec:APND}

In order to simplify the numerical calculations and to nondimensionalise the problem, we use the Szebehely convention \cite{Szebehely-book}, 
\begin{equation}\label{eq:parameters}
\mathcal{M}_1=1-\mu,\,\mathcal{M}_2=\mu, \,
x_1=-\mu,\,x_2=1-\mu,
\end{equation}
where $\mu \in [0, 1/2]$, is the only control parameter for the system, and the center of mass always lies at the origin. The dynamics of the system will be studied through the Poincar\'e sections method and the Lyapunov exponents. From now on, we set $\mu=10^{-3}$, $c=1$, and units of time such that the angular velocity of the  primaries around their common center of mass is $\omega=1$. Moreover, with the aim of observing the transition from the classical to the pseudo-Newtonian regime, we introduce the following transformation 
$$\frac{1}{c^4}\to \epsilon \frac{1}{c^4}, $$
where $\epsilon \in [0,1]$, taking the value $\epsilon=0$ in the classical limit and $\epsilon=1$ in the pseudo-Newtonian case. 

\subsection{Dynamics of the pseudo-Newtonian equations}
\begin{figure}[h!]
\centering
\begin{tabular}{ccc}
\includegraphics[width=4.0 cm,angle=0]{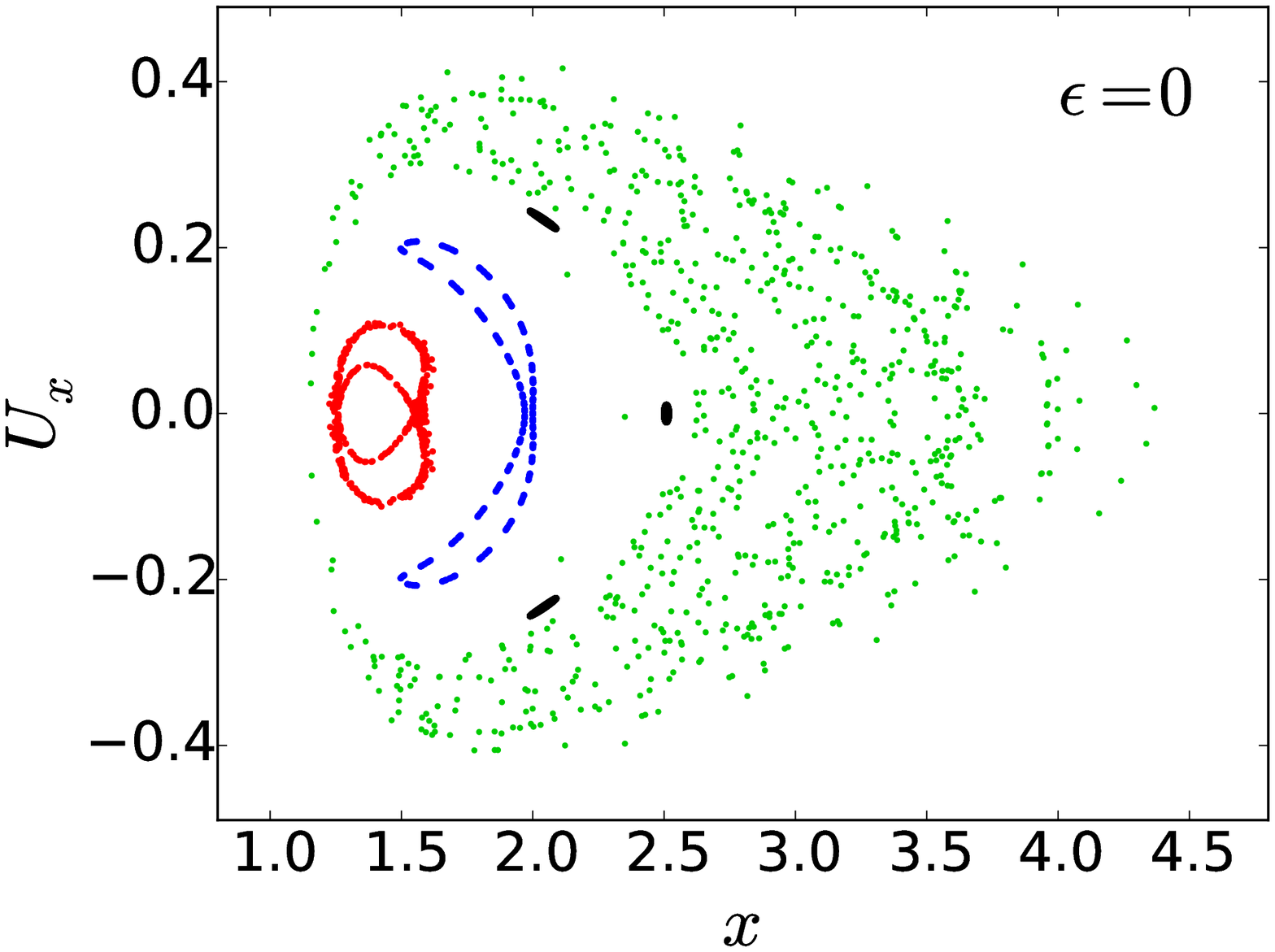}&
\includegraphics[width=4.0 cm,angle=0]{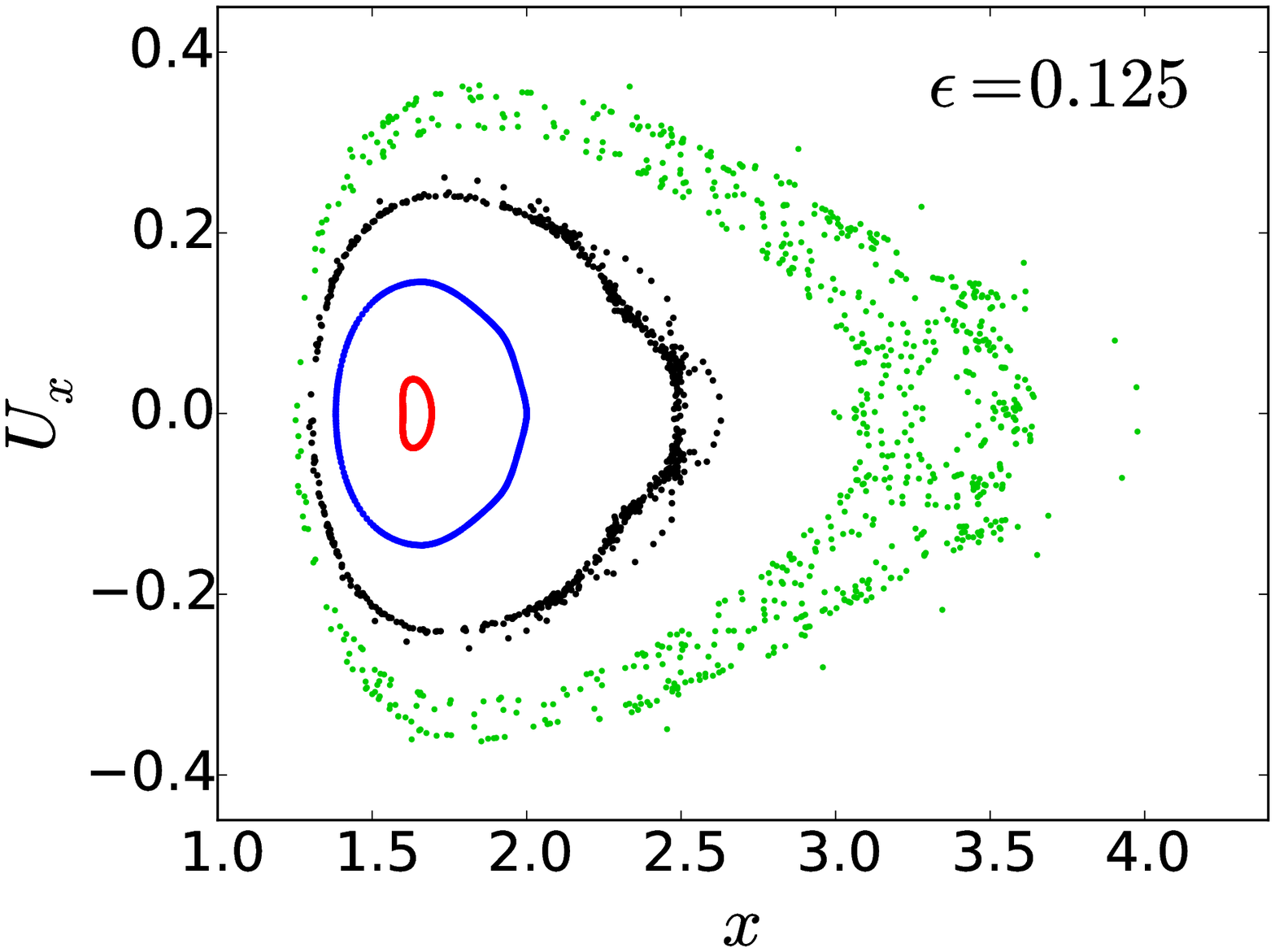}&
\includegraphics[width=4.0 cm,angle=0]{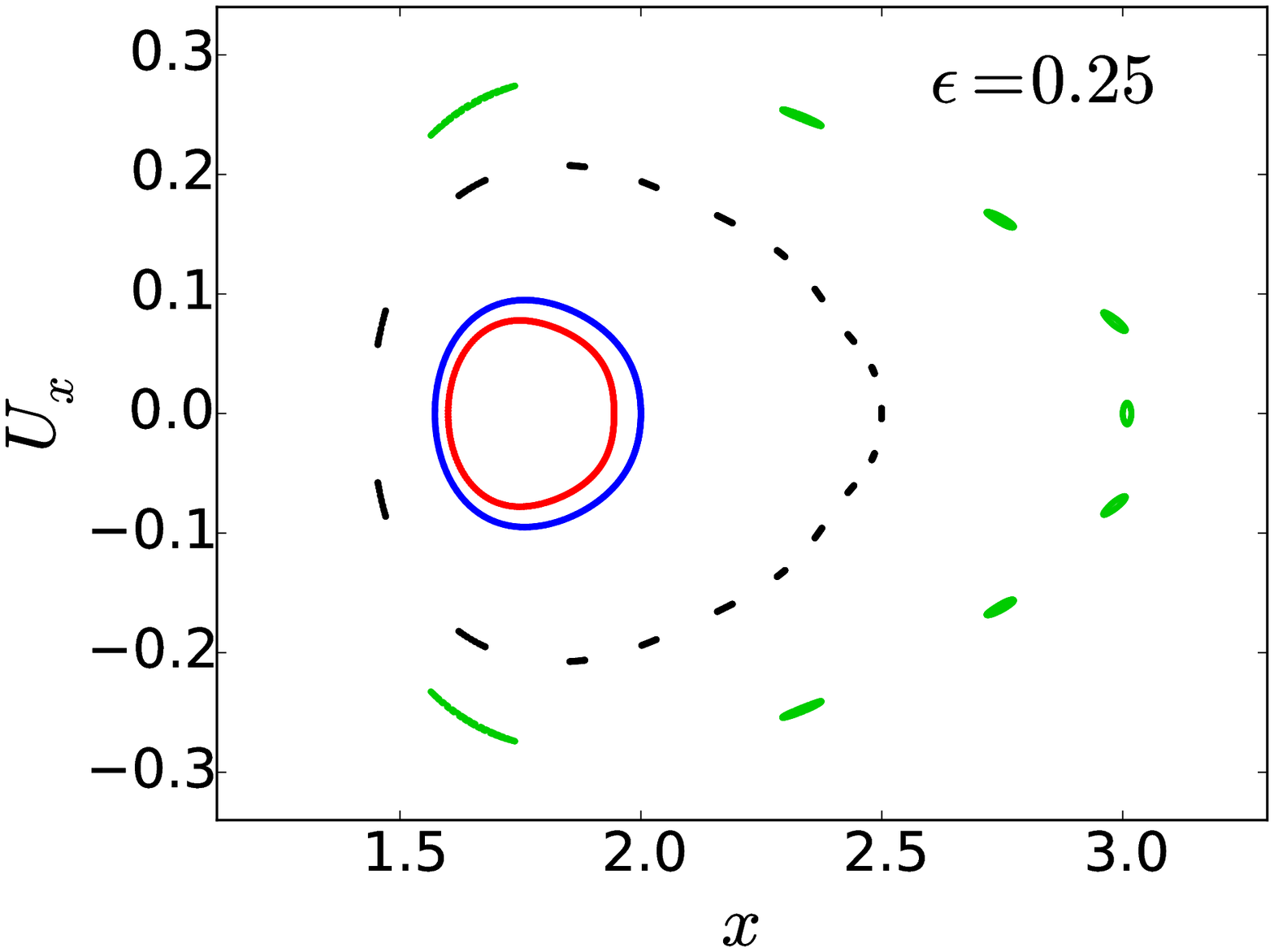}\\
\includegraphics[width=4.0 cm,angle=0]{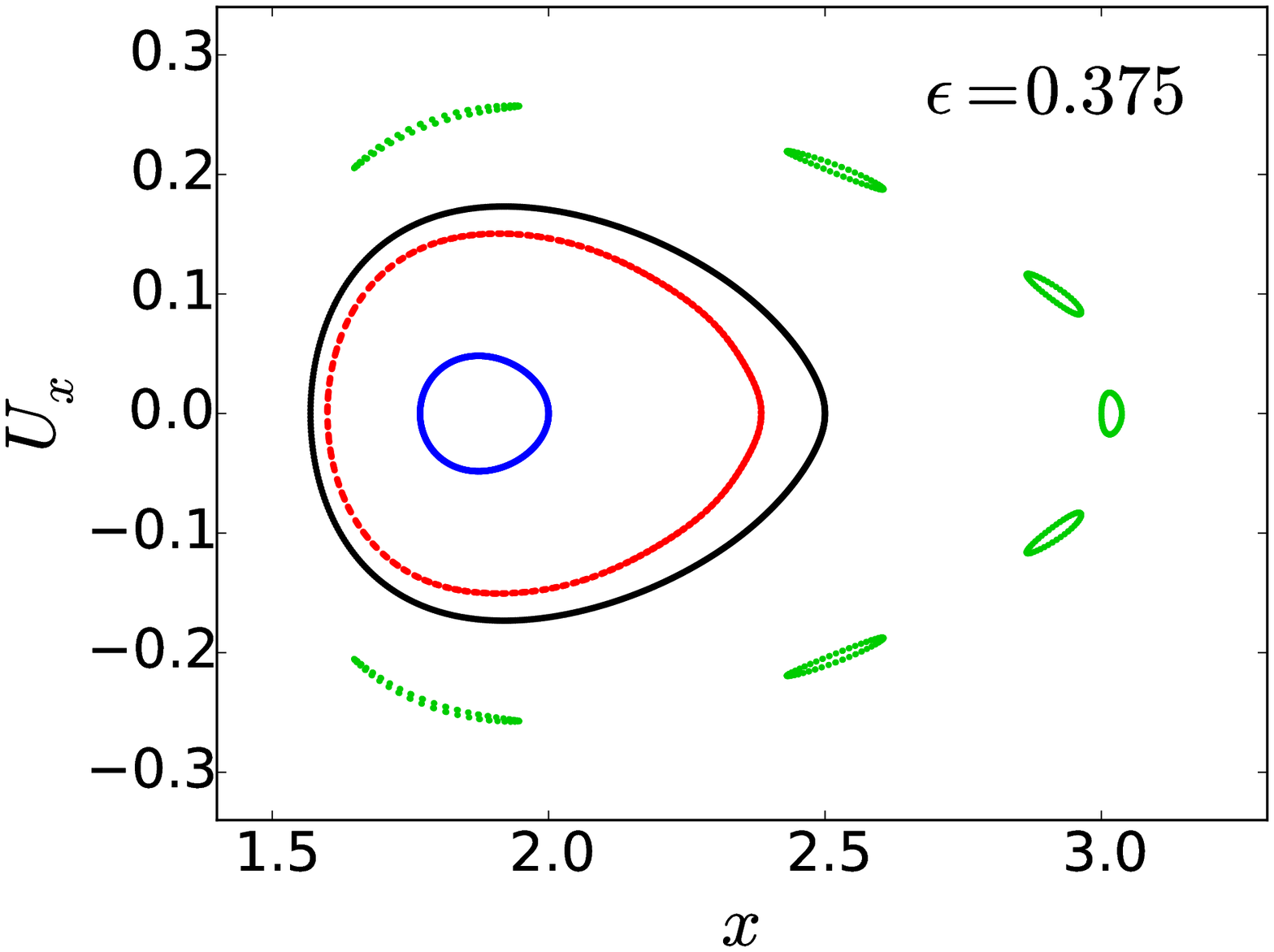}&
\includegraphics[width=4.0 cm,angle=0]{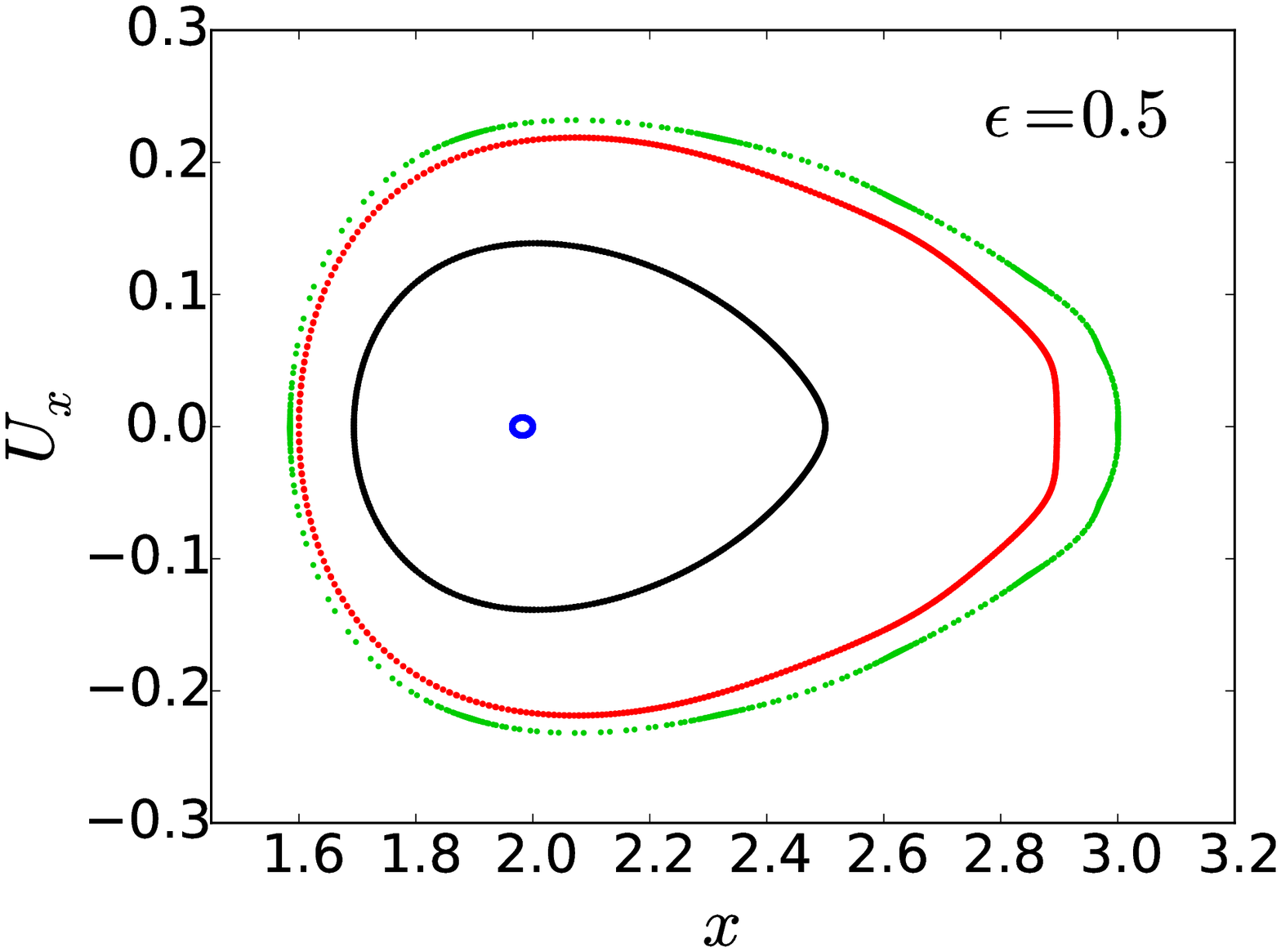}&
\includegraphics[width=4.0 cm,angle=0]{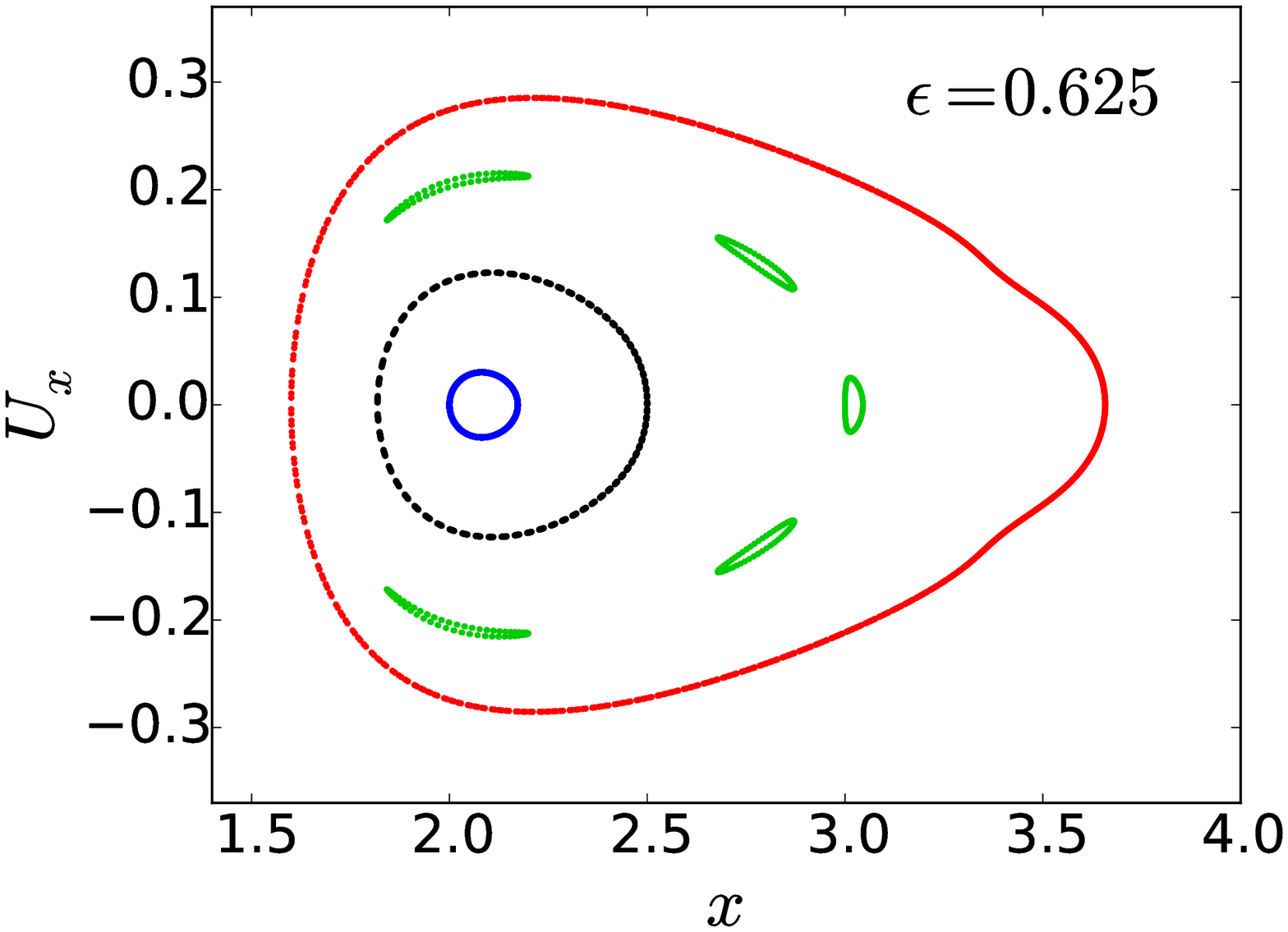}\\
\includegraphics[width=4.0 cm,angle=0]{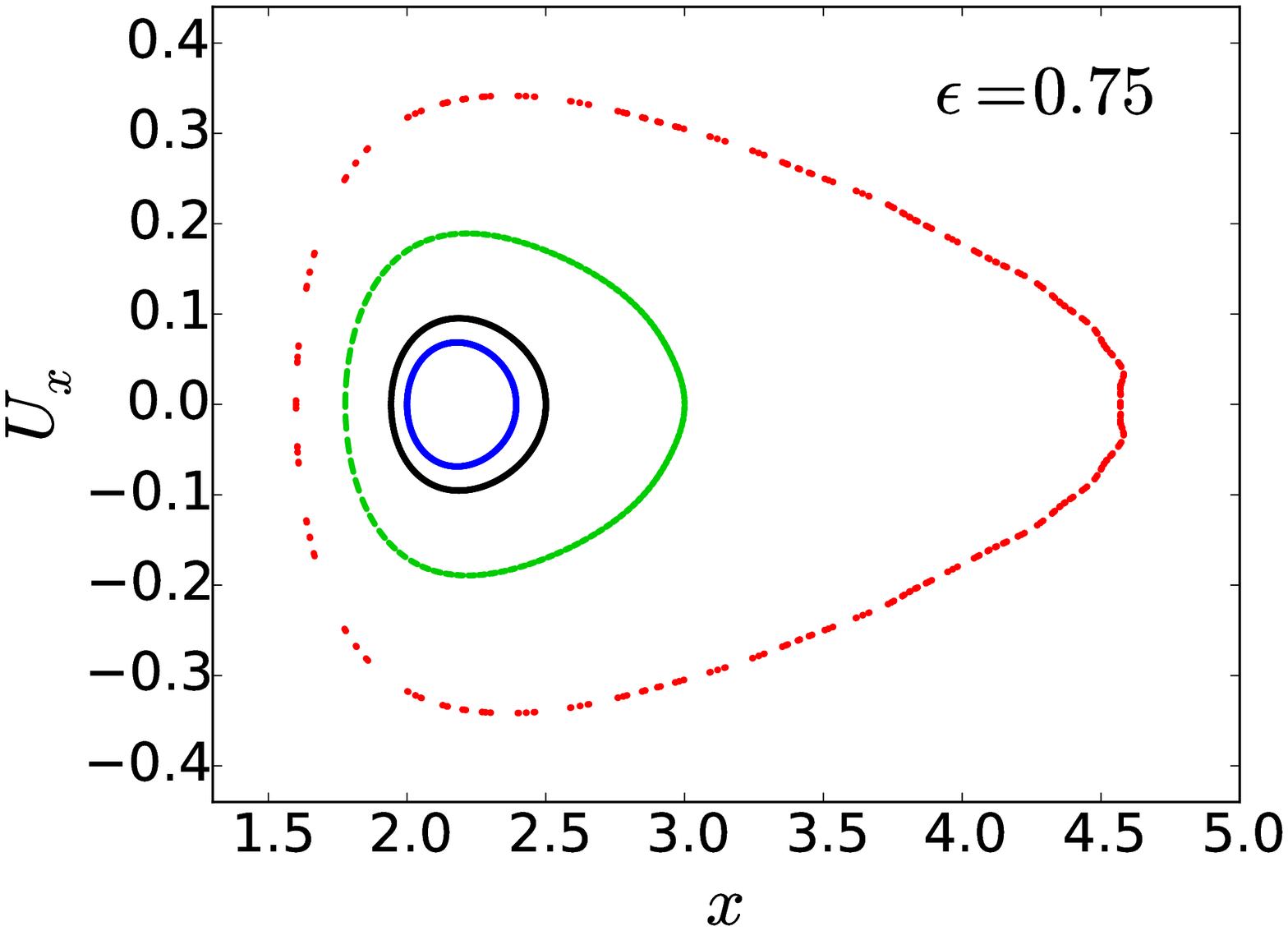}&
\includegraphics[width=4.0 cm,angle=0]{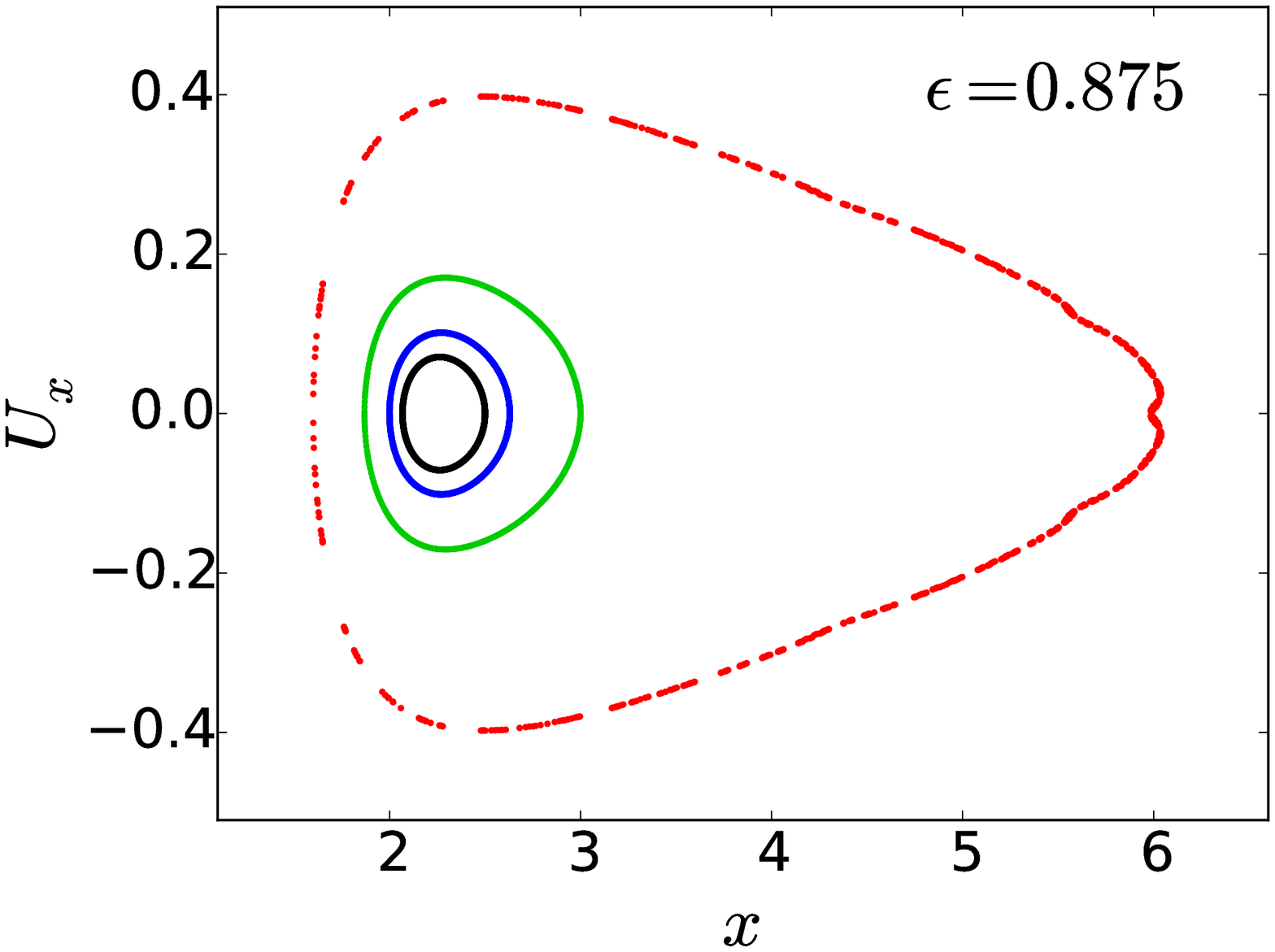}&
\includegraphics[width=4.0 cm,angle=0]{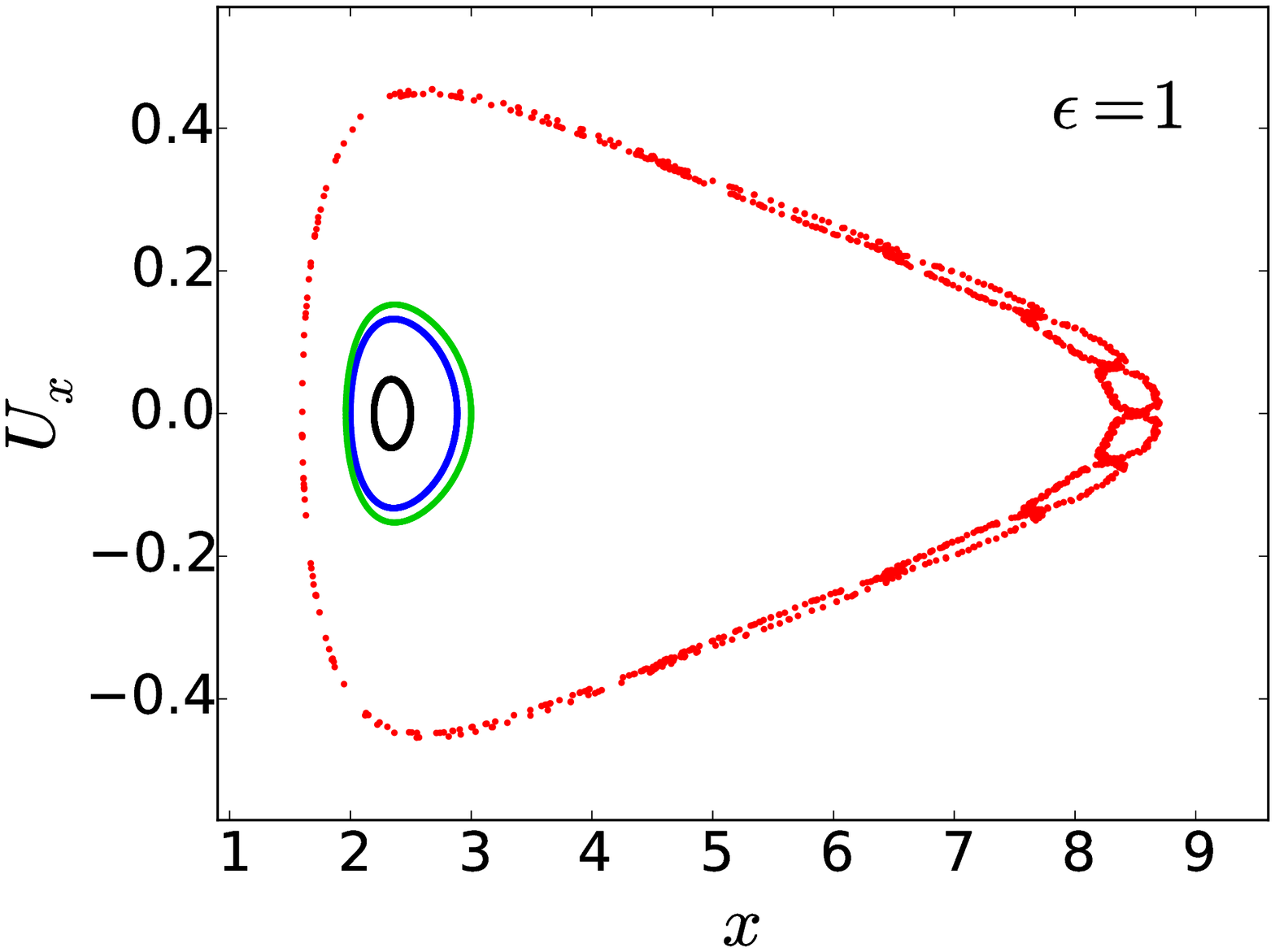}
\end{tabular}
\caption{(Color online) Poincar\'e surface of sections for the CRTBP for fixed Jacobian constant $C_{J}= 3.07$, in terms of the parameter $\epsilon$. Orbits for the sets 1 (red), 2 (blue), 3 (black), and 4 (green) have initial values $y_0=U_{x0}=0$ and $x_{0}= 1.6, 2.0, 2.5$, and $3.0$, respectively.}
\label{fig1}
\end{figure}

\begin{figure}[h!]
\centering
\includegraphics[width=7.5cm,angle=0]{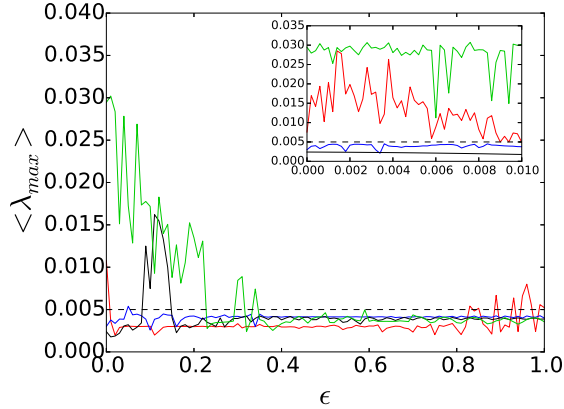}
\caption{(Color online) Average Largest Lyapunov exponent $\langle {\lambda_{max}} \rangle$ over an ensemble of $10^6$ nearby trajectories calculated with the aid of the variational method for the set of initial conditions given in Figure \ref{fig1}, for different values of $\epsilon$. The inset shows the average value of $\lambda_{max}$ for small $\epsilon$.}
\label{fig2}
\end{figure}

 In the case under consideration, we follow the evolution of the system while keeping the Jacobian constant fixed to the value $C_{J}=3.07$, see Figs. \ref{fig1} and \ref{fig2}. Given the initial conditions for $x_0, y_0$ and $U_{x0}$, the initial condition for $U_{y0}$ is determined by Eq. (\ref{eq:JacF}). It should be noted that the orbits in the Poincar\'e sections must not cross between them, because the integral of motion is the same for the set of initial conditions considered in each phase space. Orbits for the sets 1, 2, 3, and 4 have initial values $y_0=U_{x0}=0$ and $x_{0}= 1.6, 2.0, 2.5$, and $3.0$, respectively. The  convention in Fig. \ref{fig1}  and \ref{fig2} is the following: set of initial conditions 1 is plotted in red color, set 2 in blue, set 3 in black and set 4 in green. 

As an additional tool to determine the existence of chaos or regularity of the orbits, we measure the average Largest Lyapunov exponent $\langle {\lambda_{max}} \rangle$, for each trajectory (Fig. \ref{fig2}). To do so, we use the variational method (which is very accurate for many classes of dynamical systems) instead of the two-particle approach, because it has been previously shown that the last one could lead to inconsistent values of the $\lambda_{max}$, in particular when using arbitrary values of the renormalization time and the initial separation between trajectories \cite{Dubeibe2014}. The Lyapunov exponents larger than the threshold (dashed black line) can be considered chaotic, while the ones below the threshold are considered regular.

From Fig. \ref{fig1},  it can be noted that in the Newtonian problem ($\epsilon=0$) there exist chaotic orbits (green and red) for the set of initial condition 1 and 4, coexisting with regular orbits for the sets 2 and 3. Two special cases can be considered in this figure: the orbit for the set 1 (red) corresponds to chaos in a very narrow zone (weakly perturbed KAM tori) and should correspond to a small value of $\lambda_{max}$; the other case corresponds to orbit 3 (black), in which three small regular islands are present (surviving tori). If the $\epsilon$-parameter is slightly increased, the main dynamical changes are observed for this two orbits, as is expected. For values of $\epsilon$ larger than 0.4, all the sets of orbits become regular. However, in the pseudo-Newtonian limit ($\epsilon$ = 1), the narrow chaotic zone appears again for the set 1, while all the other sets keep regular. 

The Lyapunov exponent gives us a more detailed description of the system dynamics. From Fig. \ref{fig2}, the large chaotic zone, corresponding to the set of initial conditions 4, gradually reduces its chaoticity and becomes regular for larger values of $\epsilon$. A very different behavior is observed for the set of initial condition 1. In the classical limit, the orbit shows a small value of $\lambda_{max}$, for the intermediate values of $\epsilon$ the orbit turns out regular, and in the pseudo-Newtonian limit the small value of $\lambda_{max}$ emerges again. On the other hand, some sporadic appearances of chaos occur for the sets of initial conditions 2 and 3, but in the limits, $\epsilon=0$ and $\epsilon = 1$, they are regular.

\section{Concluding Remarks}\label{sec:CR}

In the present paper, we propose a new pseudo-Newtonian formulation of the planar circular restricted three-body problem, by using the Fodor-Hoenselaers-Perj\'es procedure. For this new approximation, the Jacobian constant is strictly conserved (unlike the Jacobian obtained in the first-order post-Newtonian approximation see {\it e.g} \cite{Huang2014}), and its equations of motion reduce to its classical counterpart in the limit $1/c \to 0$. Through the Poincar\'e section method and validated with the Largest Lyapunov exponent, we have shown that for the set of initial conditions and Jacobian constant selected, the Newtonian and pseudo-Newtonian CRTBP exhibit a mixed phase space, {\it i.e} regular and chaotic orbits coexist. 

The introduction of an arbitrary parameter $\epsilon$ allowed us to explore the transition from the Newtonian to the pseudo-Newtonian regime. If we track the evolution of the system keeping the Jacobian constant fixed, a chaotic orbit in the Newtonian system can be either chaotic or regular in the pseudo-Newtonian limit. In the transition, the phase space can be filled by periodic orbits even if some of the orbits are chaotic in the limits $\epsilon = 0$ and $\epsilon = 1$. In accordance with previous studies \cite{Huang2014}, in most of the  cases we found that a given set of initial conditions, whose phase space is bounded in the classical regime,  correspond to unbounded trajectories in the non-Newtonian regime, that is, the system becomes unstable. The instability of the orbits is a result of the lack of stable fixed points in the system. In fact, the number of real roots of the system, when $\ddot{x}=\ddot{y}=\dot{x}=\dot{y}=0$ in Eqs.~(\ref{ecFHPx}) and (\ref{ecFHPy}), varies when varying the mass parameter $\mu$ in the interval $[0,1/2]$.

In conclusion, we may say that even the smallest corrections to the Newtonian circular restricted three-body problem, could drastically change the dynamics of the system. In addition, it is important to note that the procedure outlined in the present paper can be used to model different kinds of sources, for example, a pair of massive spinning primaries in circular orbits, or a binary system formed by two non-spherical spinning sources, just to name a few. Results in this direction will be reported soon. 

\section*{Acknowledgments}
FLD acknowledges financial support from the University of the Llanos, under Grants Commission: Postdoctoral Fellowship Scheme. FDLC and GAG gratefully acknowledges the financial support provided by VIE-UIS, under grants numbers 1822, 1785 and 1838, and COLCIENCIAS, Colombia, under Grant No. 8840.

\end{document}